\documentclass{ifacconf}
\makeatletter
\let\old@ssect\@ssect 
\makeatother

\usepackage{natbib}
\usepackage[colorlinks=true, linkcolor=blue, citecolor=blue, urlcolor=blue]{hyperref}  
\makeatletter
\def\@ssect#1#2#3#4#5#6{%
  \NR@gettitle{#6}
  \old@ssect{#1}{#2}{#3}{#4}{#5}{#6}}

\usepackage{url}
\newtheorem{theorem}{Theorem}
\newtheorem{proposition}[theorem]{Proposition}%

\newtheorem{remark}{Remark}%
\usepackage{tikz}
\newtheorem{definition}{Definition}%

\usepackage{amsfonts,amssymb,amsmath}
\usepackage{graphicx,graphics,epsfig,color}
\usepackage{multicol}
\usepackage{algorithm}
\usepackage{algpseudocode}

\newif\ifitsdraft

\usepackage{float}
\usepackage{tikz}
\usepackage{pgfplots}
\pgfplotsset{compat=1.18}
\usepackage{comment}

\newcommand{\Inc}{\textrm{Inc}}

\newcommand{\Int}{\textrm{Int}}
\newcommand{\cl}{\textrm{cl}}

\newtheorem{assumption}{Assumption}
\newtheorem{lemma}{Lemma}
\begin{document}
\begin{frontmatter}

\title{On Robust Controlled Invariants for Continuous-time Monotone Systems} 

\thanks[footnoteinfo]{}

\author[First]{Emmanuel Jr. Wafo Wembe} 
\author[First]{Adnane Saoud} 

\address[First]{College of Computing,  University Mohammed VI Polytechnic,\\ UM6P,  Benguerir, Morocco, \\
        (email \{emmanueljunior.wafowembe,adnane.saoud\}@um6p.ma)}

\begin{abstract}                
This paper delves into the problem of computing robust controlled invariants for monotone continuous-time systems, with a specific focus on lower-closed specifications. We consider the classes of state monotone (SM) and control-state monotone (CSM) systems, we provide the structural properties of robust controlled invariants for these classes of systems and show how these classes significantly impact the computation of invariants. Additionally, we introduce a notion of feasible points, demonstrating that their existence is sufficient to characterize robust controlled invariants for the considered class of systems. The study further investigates the necessity of reducing the feasibility condition for CSM and Lipschitz systems, unveiling conditions that guide this reduction. Leveraging these insights, we construct an algorithm for the computation of robust controlled invariants. To demonstrate the practicality of our approach, we applied the developed algorithm to the coupled tank problem.  
\end{abstract}


\end{frontmatter}
\section{INTRODUCTION}
\label{sec:1}

Controlled invariants have engendered extensive literature, manifesting in both theoretical and practical applications.
Within the domain of dynamical system theory,  controlled invariants, also denoted as viable sets in \cite{aubin2009viability}, are pivotal. They are sets wherein trajectories initiated within them, subject to corresponding controls, remain within the defined set. Their applications span various problems in system analysis, notably the exploration of attractor existence, system performance, robustness, and practical stability, as shown in \cite{blanchini2008set}. Practical applications are prominent in safety-critical systems such as autonomous cars, robotics, biology, and medicine via meticulous verification of safety constraints.
A strategic imperative involves discerning a controlled invariant set characterized by a null intersection with the unsafe set, thereby ensuring the system's strict adherence to prescribed safety measures.

Linking this theoretical foundation to empirical research, \cite{blanchini2008set} relates the study of controlled invariants to Lyapunov stability. Various methodological approaches have been proposed, ranging from employing linear programming techniques by \cite{sassi2012computation} and semi-definite programming by \cite{korda2014convex} for polynomial systems, to the computation of interval-controlled invariants tailored to a general class of nonlinear-systems in \cite{saoud2021computation}. Furthermore, the exploration of symbolic control techniques in  \cite{tabuada2009verification} and \cite{saoud2019compositional} is undertaken to augment both the comprehension and computational efficiency of controlled invariant sets. This intricate scholarly landscape underscores the foundational role of controlled invariant sets in the systematic analysis and design of dynamic systems. {In \cite{dorea1999b}, authors provide necessary and sufficient conditions for invariance in polyhedral sets for linear continuous-time systems, with applications to the controllability of systems subject to trajectory constraints. The authors also provide extensions to constrained and additively disturbed systems. In \cite{DU202012571}, the authors outline conditions for the existence of a positively invariant polyhedron in positive linear systems with bounded disturbances, expressed as solvable linear programming inequalities. The paper also establishes a connection between Lyapunov stability and positively invariant polyhedra. In \cite{choi2023forward}, the authors explore the interconnections between reachability, controlled invariance, and control barrier functions. They introduce the concept of \textquotedblleft Inevitable Forward Reachable Tube\textquotedblright $\,$  as a tool to analyze controlled invariant sets and establish a strong link between these concepts. }

\textbf{Related work:} The computational investigation of controlled invariants within continuous-time systems has primarily focused on sets defined by multidimensional intervals. Within this domain, the treatment of continuous-time monotone autonomous multi-affine systems has been addressed in a seminal work by \cite{abate2009box}, with subsequent extensions to systems incorporating inputs detailed in the study by \cite{meyer2016robust}, and discrete-time monotone systems in \cite{saoud2024characterization}. Expanding upon the foundation in \cite{saoud2024characterization}, our work extends the proposed approach to continuous-time systems. In contrast to the discrete case, which focused on closed-loop robust invariance, our emphasis centers predominantly on open-loop robust invariance. We also define two less classes of monotone systems. Then, using feasible trajectories, we deduce an algorithmic procedure to compute robust controlled invariants. 

The remainder of this paper is organised as follows. Section \ref{sec:2} introduces some preliminary tools. Section \ref{sec:3} introduced the considered class of monotone systems and robust controlled invariants. Section \ref{sec:4} analyses the structural properties of robust controlled invariants for the considered class of systems, introduces the concept of feasibility, and presents an algorithmic approach to compute these invariants. Finally, Section \ref{sec:5} presents a numerical example, showing the merits of the proposed approach. 

Due to space constraints, the proofs are omitted and will
be published elsewhere.


\section{Preliminaries}
\label{sec:2}
\subsection{Notation} 
The symbols $\mathbb{N}$, $ \mathbb{N}_{>0} $, $\mathbb{R}$ and $\mathbb{R}_{>0}$ denote the set of positive integers, non-negative integers, real and non-negative real numbers, respectively. Given $n \in \mathbb{N}_{>0}$ and a set $Y \subseteq \mathbb{R}^n$, $Y^{\mathbb{R}}$denotes the set of functions from  $\mathbb{R}_{\geq 0}$ to $Y$. Given a nonempty set $K$, $\Int(K)$ denotes its interior, $\cl(K)$ denotes its closure, $\partial K$ denotes its boundary and $\overline{K}$ is its complement. The Euclidean norm is denoted by $\|.\|$. For $x \in \mathbb{R}^n$ and for $\varepsilon \geq 0$, $\mathcal{B}_{\varepsilon}(x)=\{z \in \mathbb{R}^n \mid \|z-x\| \leq \varepsilon \}$ and for a set $K \subseteq \mathbb{R}^n$, $\mathcal{B}_{\varepsilon}(K)= \cup_{x \in K}\mathcal{B}_{\varepsilon}(x)$. For $f \in Y^{\mathbb{R}}$ , $\|f\|_{\infty}$ is defined by $\|f\|_{\infty} = \sup\limits_{t \geq 0} \|f(t)\| \in \mathbb{R}_{\geq 0} \cup \{+\infty \}$. If $K$ is a closed subset of $\mathbb{B}$, we denote by $\pi_K(x) = \{y \in K | ||x -y|| = \inf\limits_{z \in K} ||z-x|| \}$. We denote by $ x_i \underset{X}{\rightarrow}x $ that for all $i\in \mathbb{N}_{\geq 0}$ , $x_i \in  X$ and $ x_i \underset{n \rightarrow \infty}{\rightarrow}x $. We denote by  $ t_i \downarrow l$  that $t_i$ is a strictly decreasing sequence that converges to $l$. For $x\in \mathbb{R}$ we define $\lfloor x\rfloor$ and $\lceil x \rceil$, the consecutive integers such that $ \lfloor x\rfloor \leq x< \lceil x \rceil$. Given a proposition $P$, $ 
\neg P$ denotes its negation.
\subsection{Partial orders}
A partially ordered set $\mathcal{L}$ has an associated binary relation $\leq_{\mathcal{L}}$ where for all $l_1,l_2,l_3 \in \mathcal{L}$, the binary relation satisfies: (i) $l_1 \leq_{\mathcal{L}} l_1 $, (ii) if $l_1 \leq_{\mathcal{L}} l_2$ and $l_2 \leq_{\mathcal{L}} l_1$ then $l_1=_{\mathcal{L}}l_2$ and, (iii) if $l_1 \leq_{\mathcal{L}} l_2$ and $l_2 \leq_{\mathcal{L}} l_3$ then $l_1 \leq_{\mathcal{L}} l_3$. If neither $l_1 \leq_{\mathcal{L}} l_2$ nor $l_2 \leq_{\mathcal{L}} l_1$ holds, we say that $l_1$ and $l_2$ are incomparable. The set of all incomparable couples in $\mathcal{L}$ is denoted by $\Inc_{\mathcal{L}}$. We say that $l_1<_{\mathcal{L}} l_2$ iff $l_1\leq_{\mathcal{L}} l_2$ and $l_1\neq_{\mathcal{L}} l_2$. Finally, a partial ordering $m \leq_{\mathcal{L}^{\mathbb{R}}} n$ between a pair of functions of a (real variable) with values in $\mathcal{L} $ holds if and only if $m(t) \leq_{\mathcal{L}} n(t) $. For a partially ordered set $\mathcal{L}$, closed intervals are $[x,y]_{\mathcal{L}}:=\{z \mid x \leq_{\mathcal{L}} z \leq_{\mathcal{L}} y\}$.
\begin{definition}
     Given a partially ordered set $\mathcal{L}$, for $a\in \mathcal{L}$, we define,  $\downarrow a :=\{x \in \mathcal{L} \mid x \leq_{\mathcal{L}} a \}$ and $\uparrow a :=\{x \in \mathcal{L} \mid a \leq_{\mathcal{L}} x \}$. When $A \subseteq \mathcal{L}$ then its lower closure (respectively upper closure) is $\downarrow A :=\bigcup_{a \in A} \downarrow a $ (respectively $\uparrow A :=\bigcup_{a \in A} \uparrow a $). A subset $A \subseteq \mathcal{L}$ is said to be {\it lower-closed} (respectively {\it upper-closed}) if $\downarrow A = A$ (respectively $\uparrow A = A$).
\end{definition}
 We have the following definitions relative to partially ordered sets.

\begin{definition}
Let $\mathcal{L}$ be a partially ordered set and $A\subseteq \mathcal{L}$. The set $A$ is said to be {\it bounded below} (in $\mathcal{L}$) if there exists a compact set $B \subseteq \mathcal{L}$ such that $A \subseteq \, \uparrow B$. Similarly, the set $A$ is said to be {\it bounded above} (in $\mathcal{L}$) if there exists a compact set $B \subseteq  \mathcal{L}$ such that $A \subseteq \,\downarrow B$.
\end{definition}

\begin{definition}
\label{def:min}
    Let $\mathcal{L}$ be a partially ordered set and consider a closed subset $A\subseteq \mathcal{L}$. If the set $A$ is bounded below then the set of {\it minimal elements} of $A$ is defined as
    $\min(A):=\{x \in A \mid \forall x_1 \in A,\, x \leq_\mathcal{L} x_1 \text{ or } (x,x_1)\in \Inc_\mathcal{L}\}$. Similarly, if the set $A$ is bounded above then the set of {\it maximal elements} of $A$ is defined as $\max(A) := \{x \in A \mid \forall x_1 \in A,\, x \geq_\mathcal{L} x_1 \text{ or } (x,x_1)\in \Inc_\mathcal{L}\}$.
\end{definition}
In the subsequent sections, our emphasis will be on lower closed sets. It is important to acknowledge that analogous results can also be achieved for upper closed sets.

\subsection{Continuous-time control systems}

In this paper, we consider the class of continuous-time control systems $\Sigma$ of the form:
\begin{equation}
\label{dis_sys}
\dot{x} = f(x,u,d)
\end{equation}
where $x \in \mathcal{X}$ is a state, $u \in \mathcal{U}$ is a control input and $d \in \mathcal{D}$ is a disturbance input. The trajectories of (\ref{dis_sys}) are denoted by $\Phi(.,x_0,\mathbf{u},\mathbf{d})$ where $\Phi(t,x_0,\mathbf{u},\mathbf{d})$ is the state reached at time $t \in \mathbb{R}_{\geq 0}$ from the initial state $x_0$ under the control input $\mathbf{u}:\mathbb{R}_{\geq 0} \rightarrow \mathcal{U}$ and the disturbance input $\mathbf{d}:\mathbb{R}_{\geq 0} \rightarrow \mathcal{D}$. For $X \subseteq \mathcal{X}$, $U \subseteq \mathcal{U}$, $D \subseteq \mathcal{D}$, and $t \geq 0$ we use the notation $\Phi(t,X,U,D) = \{\Phi(t,x, \mathbf{u},\mathbf{d}) \mid x\in X , \mathbf{u} \in U^{\mathbb{R}} , \mathbf{d} \in D^{\mathbb{R}}\}$ to denote the reachable set of the system $\Sigma$ at time $t$ from an initial condition $x \in X$ under a control input $\mathbf{u}:\mathbb{R}_{\geq 0} \rightarrow U$ and a disturbance input $\mathbf{d}:\mathbb{R}_{\geq 0} \rightarrow D$.

In the rest of the paper, we make the following assumption:

\begin{assumption}
\label{ass:1}
The system $\Sigma$ in (\ref{dis_sys}) is well-posed, i.e, for all $x \in \mathcal{X}$ , for all $\mathbf{u} \in U^{\mathbb{R}}$, for all $\mathbf{d} \in D^{\mathbb{R}}$, we have a unique solution of the system $\Sigma$, $\phi(t,x,\mathbf{u},\mathbf{d})$ defined for all $t\geq 0$. See \cite{khalil2002nonlinear}, \cite{ANGELI1999209}. 
\end{assumption}

\section{Monotones systems and preliminaries on invariance}
\label{sec:3}
\subsection{Monotones systems}
In this section, we describe classes of monotone continuous-time control systems, emphasizing their capacity to maintain order on their states and control inputs. Subsequently, we furnish characterizations of the considered classes of systems.

\begin{definition}
Consider the continuous-time control system $\Sigma$ in (\ref{dis_sys}).
The system $\Sigma$ is said to be:
\begin{itemize}
    
    \item {\it State monotone (SM)} if its sets of states and disturbance inputs are equipped with partial orders $\leq_{\mathcal{X}}$ and $\leq_{\mathcal{D}^{\mathbb{R}}}$, respectively, and for all $x_1,x_2 \in \mathcal{X}$, for all $\mathbf{u} \in \mathcal{U}^{\mathbb{R}}$ and for all $\mathbf{d}_1,\mathbf{d}_2\in \mathcal{D}^{\mathbb{R}}$, if $x_1 \leq_{\mathcal{X}} x_2$ and $\mathbf{d}_1\leq_{\mathcal{D}^{\mathbb{R}}} \mathbf{d}_2$ then $\Phi(t,x_1,\mathbf{u},\mathbf{d}_1) \leq_{\mathcal{X}} \Phi(t,x_2,\mathbf{u},\mathbf{d}_2)$, for all $t \geq 0$;
    \item {\it Control-state monotone (CSM)} if its sets of states, inputs and disturbances are equipped with partial orders, $\leq_{\mathcal{X}}$, $\leq_{\mathcal{U}^{\mathbb{R}}}$ and $\leq_{\mathcal{D}^{\mathbb{R}}}$, respectively, and for all $x_1,x_2 \in \mathcal{X},\ \mathbf{u}_1,\mathbf{u}_2 \in \mathcal{U}^{\mathbb{R}}$ and for all $\mathbf{d}_1,\mathbf{d}_2\in \mathcal{D}^{\mathbb{R}}$, if $x_1 \leq_{\mathcal{X}} x_2$, $\mathbf{u}_1 \leq_{\mathcal{U}^{\mathbb{R}}} \mathbf{u}_2$ and $\mathbf{d}_1 \leq_{\mathcal{D}^{\mathbb{R}}} \mathbf{d}_2$ then $ \Phi(t,x_1,\mathbf{u}, \mathbf{d}_1)\leq_{\mathcal{X}} \Phi(t,x_2,\mathbf{u}_2,\mathbf{d}_2)$, for all $t \geq 0$.
\end{itemize}
\end{definition}


\begin{remark}
{In the previous definitions, we suppose monotonicity with respect to disturbance inputs. The above definitions indicate that a CSM system is a SM system. The SM, CSM properties defined in this paper correspond to the concept of monotonicity in~\cite{smith2008monotone} and~\cite{meyer2015adhs}, respectively.}
\end{remark}

\begin{remark}
    In practice, the monotonicity of a system can be verified using other characterizations (see Appendix \ref{sec:charMono}).
\end{remark}







Next, we first introduce an auxiliary result, enabling us to present equivalent characterizations of the proposed classes of monotone systems.

\begin{proposition}
\label{prop:characterizations_monotone}
Consider the control system $\Sigma$ in (\ref{dis_sys}). We have the following properties:
\begin{itemize}
    \item[(i)] The system $\Sigma$ is SM if and only if for all $x \in X$, $\mathbf{u} \in U^{\mathbb{R}} $ and $\mathbf{d} \in D^{\mathbb{R}}$ we have for all $t \geq 0$
    $$ \Phi(t,\downarrow x,  \mathbf{u}, \downarrow \mathbf{d}) \subseteq \downarrow \Phi(t, x,\mathbf{u}, \mathbf{d})$$
    \item[(ii)] The system $\Sigma$ is CSM if and only if for all $x \in X$, $\mathbf{u} \in U^{\mathbb{R}} $ and $\mathbf{d} \in D^{\mathbb{R}}$ we have for all $t \geq 0$
    $$ \Phi(t,\downarrow x,  \downarrow \mathbf{u}, \downarrow \mathbf{d}) \subseteq \downarrow \Phi(t, x,\mathbf{u}, \mathbf{d})$$
\end{itemize}
\end{proposition}

\subsection{Preliminaries on invariance}
We introduce the concept of  robust controlled invariant for continuous-time dynamical systems consistent with \cite{dorea1999b} and \cite{korda2014convex}.

\begin{definition}
\label{def:contr_inv2}
Consider the system $\Sigma$ in (\ref{dis_sys}) and let $X \subseteq \mathcal{X}$, $U \subseteq \mathcal{U}$ and $D \subseteq \mathcal{D}$ be the constraints sets on the states, inputs and disturbances, respectively. The set $K \subseteq \mathcal{X}$ is a robust controlled invariant for the system $\Sigma$ and constraint set $(X,U,D)$ if $K \subseteq X$ and the following holds:
\begin{equation}
\label{eqn:prop_charact}
    \forall x \in K,~ \exists \mathbf{u}\in U^{\mathbb{R}}~ \text{ s.t }~ \Phi(t,x,\mathbf{u},D) \subseteq K, \, \forall t\geq 0
\end{equation}
\end{definition}


\textcolor{blue}{One major difference between open-loop and closed is the dependence of control inputs. While in an open loop, the control input depends only on the initial state, in a closed loop, the control depends on the whole trajectory.}

 From Definition~\ref{def:contr_inv2}, one can readily see that the robust controlled invariance property is closed under union. Hence, we have the existence of a unique maximal robust controlled invariant that contains all  the robust controlled invariants. 

\begin{definition}
\label{def:contr_invar}
Consider the system $\Sigma$ in (\ref{dis_sys}) and let $X \subseteq \mathcal{X}$, $U \subseteq \mathcal{U}$ and $D \subseteq \mathcal{D}$ be the constraints sets on the states, inputs and disturbances, respectively. The set $K \subseteq \mathcal{X}$ is the \emph{maximal} robust controlled invariant set for the system $\Sigma$ and constraint set $(X,U,D)$ if:
\begin{itemize}
    \item $K \subseteq \mathcal{X}$ is a robust controlled invariant for the system $\Sigma$ and constraint set $(X,U,D)$;
    \item $K$ contains any robust controlled invariant for the system $\Sigma$ and constraint set $(X,U,D)$.
\end{itemize}
\end{definition}

Enlarging the control input set and reducing the disturbance set preserve the invariance property. 

\begin{lemma}
\label{lem:inv_cont_dist}
Consider the system $\Sigma$ in (\ref{dis_sys}) and let $X \subseteq \mathcal{X}$, $U_1, U_2 \subseteq \mathcal{U}$ and $D_1, D_2 \subseteq \mathcal{D}$ be constraints sets on the states, inputs and disturbances, respectively, satisfying $U_1 \subseteq U_2$ and $D_2 \subseteq D_1$. If $K$ is a robust controlled invariant for the system $\Sigma$ and constraint set $(X, U_1, D_1)$ then $K$ is a robust controlled invariant for the system $\Sigma$ and constraint set $(X, U_2, D_2)$.
\end{lemma}


\section{Structural properties and computation of robust Controlled invariants}
\label{sec:4}
\subsection{Structural properties of robust controlled invariants}
\label{sec:4.1}
The following result introduces topological characterisations of controlled invariants for more regular systems.  

\begin{proposition}
\label{prop:closedness}
Consider the system $\Sigma$ in (\ref{dis_sys}) and let $X \subseteq \mathcal{X}$, $U \subseteq \mathcal{U}$ and $D \subseteq \mathcal{D}$ be constraints sets on the states, inputs, and disturbances, respectively. Suppose that the set of state $X$ is closed, and the set of control inputs $U$ and disturbance inputs $D$ are compact. Suppose that the dynamics $f:\mathcal{X}\times \mathcal{U} \times \mathcal{D} \rightarrow \mathcal{X}$ of the system $\Sigma$ is Lipschitz over its first argument\footnote{The map $f:\mathcal{X}\times \mathcal{U} \times \mathcal{D} \rightarrow \mathcal{X}$ is Lipschitz continuous over its first argument if there exists a constant $\lambda \geq 0$ such that for all $x_1,x_2 \in \mathcal{X}$, for all $u \in U$ and for all $d \in D$, we have $
    \|f(x_1,u,d)-f(x_2,u,d)\| \leq \lambda \|x_1-x_2\|$.}, continuous over its second and third arguments. Then the following properties hold:
\begin{itemize}
    \item[(i)] If the set $K \subseteq X$,   is a robust controlled invariant for the system $\Sigma$ and constraint set $(X,U,D)$, then the set $\cl(K)$ is a robust controlled invariant for the system $\Sigma$ and constraint set $(X,U,D)$;
    \item[(ii)] If the set $K \subseteq X$ is the maximal robust controlled invariant for the system $\Sigma$ and constraint set $(X,U,D)$, then the set $K$ is closed.
\end{itemize}
\end{proposition}

In the following, we provide different characterizations of robust controlled invariants when dealing with monotone dynamical systems and lower-closed safety specifications (i.e. a lower closed set of constraints $\mathcal{X}$ on the state-space). 

\begin{theorem}
\label{thm1}

Consider the system $\Sigma$ in (\ref{dis_sys}) and let $X \subseteq \mathcal{X}$, $U \subseteq \mathcal{U}$ and $D \subseteq \mathcal{D}$ be the constraints sets on the states, inputs and disturbances, respectively, where the set $X$ is lower closed. The following properties hold:
\begin{itemize}
    \item[(i)] If the system $\Sigma$ is SM and if a set $K$ is a robust controlled invariant of the system $\Sigma$ and constraint set $(X, U, D)$, then its lower closure is also a robust controlled invariant for the system $\Sigma$ and constraint set $(X, U, D)$;
    
    \item[(ii)] If the system $\Sigma$ is SM then the maximal robust controlled invariant $K$ for the system $\Sigma$ and constraint set $(X,U,D)$ is lower closed;

    \item[(iii)] If the system $\Sigma$ is SM and the set of disturbance inputs $D$ is closed and bounded above then the maximal robust controlled invariant for the system $\Sigma$ and constraint set $(X, U,D)$ is the maximal robust controlled invariant for the system $\Sigma$ and the constraint set $(X, U,D_{\max})$, where $D_{\max}=\max(D)$;

    \item[(iv)] If the system $\Sigma$ is CSM, the set of control inputs $U$ is closed and bounded below then the maximal robust controlled invariant for the system $\Sigma$ and constraint set $(X, U, D)$ is the maximal robust controlled invariant for the system $\Sigma$ and the and constraint set $(X, U_{\min}, D)$, where $U_{\min}=\min(U)$;
    \item[(v)]
    If the system $\Sigma$ is CSM, the set of control inputs $U$ is closed and bounded below and the set of disturbance inputs $D$ is closed and bounded above, then the maximal robust controlled invariant for the system $\Sigma$ and constraint set $(X, U,D)$ is the maximal robust controlled invariant for the system $\Sigma$ and constraint set $(X, U_{\min}, D_{\max})$, where $U_{\min}=\min(U)$ and $D_{\max}=\max(D)$.
\end{itemize}
\end{theorem}

Derived from conditions (ii) and (iii), maximal robust controlled invariants can be characterized through the utilization of maximal elements of the state space $X$, and maximal disturbance elements in $D_{\max}$. Conversely, for CSM systems detailed in (iv) , the invariance property can be ensured using only the minimal control elements in $U_{\min}$. Let us note that since the maximal robust controlled invariant set is lower closed, the boundary of this set can be taught as a Pareto Front. As such, Multidimensional binary search techniques can be applied to compute this set as in \cite{legriel2010approximating} (see section 4.3).

\begin{proposition}
\label{prop:charac2}
Consider the system $\Sigma$ in (\ref{dis_sys}) and let $X \subseteq \mathcal{X}$, $U \subseteq \mathcal{U}$ and $D \subseteq \mathcal{D}$ be the constraints sets on the states, inputs and disturbances, respectively, where the set $X$ is lower closed. Consider a closed and lower closed set $K \subseteq X$. The following properties hold:
\begin{itemize}
    \item[(i)] If the system $\Sigma$ is SM and the set of disturbance inputs $D$ is closed and bounded above then the set $K$ is a robust controlled invariant of the system $\Sigma$ and constraint set $(X,U,D)$, if and only if $ \forall x \in \max(K)$ there exist 
   $\mathbf{u} \in U ^ {\mathbb{R}}$ , such that   for any disturbance input $\mathbf{d}:\mathbb{R}_{\geq 0} \rightarrow D_{\max}$, the solution of the open-loop system $\Phi(.,x,\mathbf{u},\mathbf{d}):\mathbb{R}_{\geq 0} \rightarrow \mathcal{X}$ satisfies \begin{equation}
      \Phi(t,x,\mathbf{u},\mathbf{d}) \in K, ~  \forall t \geq 0 \label{eqn:prop_charact1}
  \end{equation}
    where $D_{\max}=\max(D)$;
    \item[(ii)] If the system $\Sigma$ is CSM, the set of control inputs $U$ is closed and bounded below and the set of disturbance inputs $D$ is closed and bounded above then the set $K$ is a robust controlled invariant of the system $\Sigma$ and constraint set $(X, U, D)$, if and only if $ \forall x \in \max(K)$ there exist 
$\mathbf{u}: \mathbb{R}_{\geq 0} \rightarrow U_{\min}$,  such that for any disturbance input $\mathbf{d}:\mathbb{R}_{\geq 0} \rightarrow D_{\max}$, the solution of the open-loop system $\Phi(.,x,\mathbf{u},\mathbf{d}):\mathbb{R}_{\geq 0} \rightarrow \mathcal{X}$ satisfies 
\begin{equation}
      \Phi(t,x,\mathbf{u},\mathbf{d}) \in K, ~  \forall t \geq 0 \label{eqn:prop_charact2}
  \end{equation}
    where $U_{\min}=\min(U)$ and $D_{\max}=\max(D)$.
\end{itemize}
\end{proposition}

Proposition~\ref{prop:charac2} introduces significant simplifications to verifying whether a lower closed set qualifies as a robust controlled invariant. Instead of examining the robust controlled invariance condition (see equation (\ref {eqn:prop_charact})) for all elements $x \in K$, $\mathbf{u} \in U^{\mathbb{R}}$, and $\mathbf{d} \in D^{\mathbb{R}}$ in the context of general nonlinear systems, the verification process can be confined to the following cases:

\begin{itemize}
    \item $x \in \max(K)$, $\mathbf{u} \in U^{\mathbb{R}}$, and $\mathbf{d} \in D_{\max}^{\mathbb{R}}$ for SM systems;
    \item $x \in \max(K)$, $\mathbf{u} \in U_{\min}^{\mathbb{R}}$, and $\mathbf{d} \in D_{\max}^{\mathbb{R}}$ for CSM systems.
\end{itemize}

Additionally, this property proves particularly advantageous in practical applications, especially when $\max(K)$ is finite, $D_{\max}$ and $U_{\min}$ are singletons, while $K$, $D$, and $U$ are infinite.

The preceding characterizations of robust controlled invariants are global criteria. While exact, these criteria pose challenges in practical verification. In the following section, using the notion of feasibility, we introduce a characterization of Robust controlled invariants.

\subsection{Feasibility and robust controlled invariance}

{In this section, We introduce the concept of feasibility. This Idea has been introduced for discrete systems in \cite{saoud2024characterization, sadraddini2018formal}.}

\begin{definition}
\label{def:feas}
 A point $x_0 \in X$ is said to be {\it feasible} with respect to the constraint set $(X,U,D)$ if there exists an input trajectory $\mathbf{u}:\mathbb{R}_{\geq 0} \rightarrow \mathcal{U}$ and $T>0$ such that 
\begin{equation}
\label{eqn:feas1o}
    \Phi(t,x_0,\mathbf{u},D) \subseteq X, \quad \forall~ 0\leq t<T 
\end{equation}
and 
\begin{equation}
\label{eqn:feas2o}
 \Phi(T,x_0,\mathbf{u},D) \subseteq \downarrow \bigcup\limits_{0 \leq  t <T} \Phi(t,x_0,\mathbf{u},D).
\end{equation}
\end{definition}

In the following, we characterize the concept of feasibility for SM systems. 

First we introduce the following characterisation of the reachable set for Monotones systems.

\begin{lemma}
\label{lem:DSM_feas}
Consider the system $\Sigma$ in (\ref{dis_sys}). If the system $\Sigma$ is SM and the set of disturbance inputs $D$ is closed and bounded above then for any point $x_0 \in X$ and any input trajectory $\mathbf{u}:\mathbb{R}_{\geq 0} \rightarrow \mathcal{U}$, we have $\Phi(t,x_0,\mathbf{u},D) \subseteq \downarrow \Phi(t,x_0,\mathbf{u},D_{\max})$, for all $t \in \mathbb{R}_{\geq 0}$, where $D_{\max}=\max(D)$.
\end{lemma}

This lemma is a direct consequence of proposition \ref{prop:characterizations_monotone}.


\begin{proposition}
\label{prop:feas}
Consider the system $\Sigma$ in (\ref{dis_sys}) and let $X \subseteq \mathcal{X}$, $U \subseteq \mathcal{U}$ and $D \subseteq \mathcal{D}$ be the constraints sets on the states, inputs and disturbances, respectively. If the system $\Sigma$ is SM, the set of states $X$ is lower closed and the set of disturbance inputs $D$ is closed and bounded above, then a point $x_0 \in X$ is feasible w.r.t the constraint set $(X, U,D_{\max})$ if and only if it is feasible w.r.t the constraint set $(X, U,D)$, where $D_{\max}=\max(D)$.
\end{proposition}

Intuitively, the result of Proposition \ref{prop:feas} shows that to check if a point is feasible, it is sufficient to explore the trajectories with the maximal disturbance inputs in the set $D_{\max}$. 

We now show how the existence of feasible trajectories makes it possible to construct robust controlled invariants.

\begin{theorem}
\label{prop:feas_inv}
Consider the system $\Sigma$ in (\ref{dis_sys}) and let $X \subseteq \mathcal{X}$, $U \subseteq \mathcal{U}$ and $D \subseteq \mathcal{D}$ be the constraints sets on the states, inputs and disturbances, respectively, where the set $X$ is lower closed. Assume that the set of disturbances $D$ is a multidimensional interval of the form $\left[d_{\min}, d_{\max}\right]$, with $d_{\min},d_{\max} \in D$. If the system $\Sigma$ is SM, then the following holds:
 If $x_0 \in X$ is feasible w.r.t the constraint set $(X,U,D)$, then there exists an input trajectory $\mathbf{u}:\mathbb{R}_{\geq 0} \rightarrow \mathcal{U}$ and $T>0$ such that the set  \begin{equation} K=\downarrow \bigcup\limits_{0 \leq t \leq T} \Phi(t,x_0,\mathbf{u},D)    \end{equation} is a robust controlled invariant for the system $\Sigma$ and constraint set $(X,U,D)$.
\end{theorem}

We also have the following characterization of feasibility for a particular class of CSM systems.

\begin{proposition}
\label{prop:feas_CSM}
Consider the system $\Sigma$ in (\ref{dis_sys}) and let $X \subseteq \mathcal{X}$, $U \subseteq \mathcal{U}$ and $D \subseteq \mathcal{D}$ be the constraints sets on the states, inputs and disturbances, respectively, where $X$ is lower closed. If the system $\Sigma$ is CSM, the set of states $X$ is lower closed, the set of control inputs $U$ is closed and bounded below, the set of disturbances $D$ is a multidimensional interval of the form $\left[d_{\min}, d_{\max}\right]$, with $d_{\min},d_{\max} \in D$ , and for all $\varepsilon \in \mathbb{R}^n_{\geq 0}$, for all $x_1,x_2 \in \mathcal{X}$ and for all $u \in U$, the following condition is satisfied:
\begin{equation}
    \label{eqn:feas_CSM_C}
    x_1 \geq x_2 +\varepsilon \implies \mathcal{B}_{\varepsilon}(\Phi(t,x_2,u,D))\subseteq \downarrow \Phi(t,x_1,u,D), \forall t>0
\end{equation}
then a point $x_0 \in X$ is feasible w.r.t the constraint set $(X,U,D)$ if and only if it is feasible w.r.t the constraint set $(X,U_{\min},D)$, where $U_{\min}=\min(U)$.
\end{proposition}

Moreover, we have the following result, characterizing a special case of feasibility for the particular class of monotone systems with Lipschitz dynamics.

\begin{theorem}
\label{thm:stric_feas}
Consider the SM system $\Sigma$ in (\ref{dis_sys}) and let $X \subseteq \mathcal{X}$, $U \subseteq \mathcal{U}$ and $D \subseteq \mathcal{D}$ be the constraints sets on the states, inputs and disturbances, respectively. Assume that the map $f: \mathcal{X} \times \mathcal{U} \times \mathcal{D} \rightarrow \mathcal{X}$ defining the system $\Sigma$ is  Lipschitz on its first argument, uniformly continuous on its second and third arguments, and the sets of control inputs $U$ and disturbance inputs $D$ are compact. For $x_0 \in \mathcal{X}$, if the following conditions are satisfied:
\begin{itemize}
    \item[(i)] $x_0$ is feasible w.r.t the constraint set $(X,U,D)$ and there exists $\mathbf{u}:\mathbb{R}_{\geq 0} \rightarrow \mathcal{U}$, $T >0$ and $\varepsilon_T$ such that 
\begin{equation}
\label{eqn:thm}
 \mathcal{B}_{\varepsilon_{T}}(\Phi(T,x_0,\mathbf{u},D)) \subseteq \downarrow \bigcup\limits_{0 \leq t <T} \Phi(t,x_0,\mathbf{u},D).
\end{equation}
    \item[(ii)] there exists $\gamma >0$ such that $ \mathcal{B}_{\gamma}(\Phi(t,x_0,\mathbf{u},D)) \subseteq X, \quad \forall~0<t<T$
\end{itemize}
 then there exists $\beta>0$ such that for any $x_1 \in \{\uparrow x_0\} \cap \mathcal{B}_{\beta}(x_0)$, $x_1$ is feasible w.r.t the constraint set $(X,U,D)$. Moreover one can explicitly determine the value of $\beta$ as a function of the parameters $\varepsilon_T$ and $\gamma$. 
\end{theorem}


In Theorem~\ref{thm1}, we examined the structural properties of the maximal robust controlled invariant set for monotone systems and lower-closed safety specifications. Additionally, in Theorem \ref{prop:feas_inv}, we derived a robust controlled invariant set from a feasible trajectory. Although the characterizations in Equations (\ref{eqn:feas1o}) and (\ref{eqn:feas2o}) are less restrictive compared to Equation (\ref{eqn:prop_charact}), the associated search space remains extensive. 

Proposition \ref{prop:feas_CSM} introduces a noteworthy reduction for CSM systems. In cases where the system tends to amplify the difference between initial conditions, the feasibility check for minimal input elements alone is sufficient. For systems exhibiting greater regularity in their dynamics, a slight modification of the feasibility condition ensures that points near a feasible point also remain feasible, as shown in Theorem \ref{thm:stric_feas}. These reductions alleviate the complexity associated with the search for feasible points within a safe set. 
In Theorems \ref{thm1} and \ref{prop:feas_inv}, we derived useful characterisations of robust controlled invariant. Proposition \ref{prop:feas_CSM} introduce conditions for checking feasibility for CSM systems using only minimal control inputs. Theorem  \ref{thm:stric_feas} show that poinst near feasible points can also exhibit feasibility property.
\subsection{Computation of Robust Controlled Invariants}

This section introduces an algorithm for computing maximal robust controlled invariants, applicable to systems  as described in (\ref{dis_sys}). The algorithm operates within the constraints defined by the set $(X,U,D)$, where $X$ is a lower-closed constraint set, and $D$ is a multidimensional interval $D=[d_{min}, d_{max}]$.

The algorithm, presented as Algorithm \ref{alg:cap}, is designed to handle both SM and CSM systems. A two-step process is employed, where the feasibility conditions for SM systems are initially considered, and then further simplifications are introduced for CSM systems.

The algorithm begins by initializing two sets, $\mathcal{F}_1$ and $\mathcal{F}_2$, as empty sets. The set $\mathcal{F}_1$ collects the states that belongs to the maximal robust controlled invariant set while the set $\mathcal{F}_2$ collects the remaining states. It iterates over the maximum and minimum elements of the lower-closed set $X$. For each element, the algorithm checks the feasibility using the command $feasible$ (line 3) by verifying conditions (\ref{eqn:feas1o}) and (\ref{eqn:feas2o}). If the conditions are met, the resulting set $Z$ from Theorem \ref{prop:feas_inv} and defined as follows:
\begin{equation}
\label{eqn:feas_set}
    Z = \downarrow \bigcup\limits_{0\leq t\leq T} \Phi(t, x, u, D_{max}) 
\end{equation}
is included in $\mathcal{F}_1$. Conversely, if the state $x$ is deemed unsafe, the set $H$ defined by
\begin{equation}
\label{eqn:feas_unsaf}
    H = \uparrow \bigcup\limits_{0\leq t\leq T} \Phi(t, x, u, D_{max}) 
\end{equation}
is added to $\mathcal{F}_2$.

The iterative refinement process then follows, where the algorithm refines $\mathcal{F}_1$ and $\mathcal{F}_2$ until their Hausdorff distance is below a predefined threshold $\epsilon$. This process involves selecting elements from the complement of both sets and checking their feasibility or unsafety. The resulting refined $\mathcal{F}_1$ is returned as an approximation of the maximal robust controlled invariant set.

\begin{algorithm}
\caption{Computation of the maximal robust controlled invariant set}\label{alg:cap}
\begin{algorithmic}[1]
\Require A system $\Sigma$ as in (\ref{dis_sys}), a constraint set $(X,U,D)$ where $X$ is lower closed and $D$ is a multidimensional interval $[d_{min}, d_{max}]$.
\Ensure $K$ an approximation of the maximal robust controlled invariant set 
\State $ \mathcal{F}_1 = \mathcal{F}_2 = \emptyset$
\For{$x \in  \max(X)$}
    \If{feasible($x$)}
        \State $\mathcal{F}_1= \mathcal{F}_1 \bigcup Z$ \Comment{$Z$ is defined in (\ref{eqn:feas_set})}
    \ElsIf{unsafe($x$)}
        \State $\mathcal{F}_2 = \mathcal{F}_2 \bigcup H$ \Comment{$H$ is defined in (\ref{eqn:feas_unsaf})}
    \EndIf
\EndFor
\If{$\mathcal{F}_1 = X$}
    \State \Return $K = \mathcal{F}_1$
\EndIf
\For{$x \in  \min(X)$}
    \If{feasible($x$)}
        \State $\mathcal{F}_1= \mathcal{F}_1 \bigcup Z$
    \ElsIf{unsafe($x$)}
        \State $\mathcal{F}_2 = \mathcal{F}_2 \bigcup H$
    \EndIf
\EndFor
\If{$\min(X) \subseteq \mathcal{F}_2$}
    \State \Return $K = \emptyset$
\EndIf

\While{$d(\mathcal{F}_1,\mathcal{F}_2) > \epsilon$}\Comment{Hausdorf distance}
    \State Select $x \in (X \setminus \mathcal{F}_2) \cap (X \setminus \mathcal{F}_1)$
    \If{feasible(x)}
        \State $\mathcal{F}_1= \mathcal{F}_1 \bigcup Z$
    \ElsIf{unsafe($x$)}
        \State $\mathcal{F}_2 = \mathcal{F}_2 \bigcup H$
    \EndIf
\EndWhile

\State \Return  $K= \mathcal{F}_1$
\end{algorithmic}
\end{algorithm}

Moreover, the performance of the algorithm can be improved by considering the two following properties:
\begin{itemize}
    \item If the system is CSM and conditions from Proposition \ref{prop:feas_CSM} are satisfied, we can limit ourselves to the constraint set $(X,U_{min}, D_{max})$, in this case, the trajectories with $u\in U_{\min}$ will first be explored.
    \item If the system $\Sigma$ is Lipschitz and satisfies conditions from Theorem \ref{thm:stric_feas}, then we can modify the set $Z$ in (\ref{eqn:feas_set}) to $Z = \downarrow \bigcup\limits_{0\leq t\leq T} \Phi(t, x, u, D_{max})  \bigcup \{\{\uparrow x\}\bigcap \mathcal{B}_{\beta}(x)\}$ with $\beta$ as defined in theorem \ref{thm:stric_feas}
\end{itemize}

\section{Numerical Applications}
\label{sec:5}

In this section, we consider the non-linear model of coupled tanks described in \cite{ALACoupledTanks}. The system is
shown in figure \ref{fig:coupl_tanks} and 
described by the following differential equation: 

\begin{figure}[h]
    \centering
    \includegraphics[width= 0.4\textwidth]{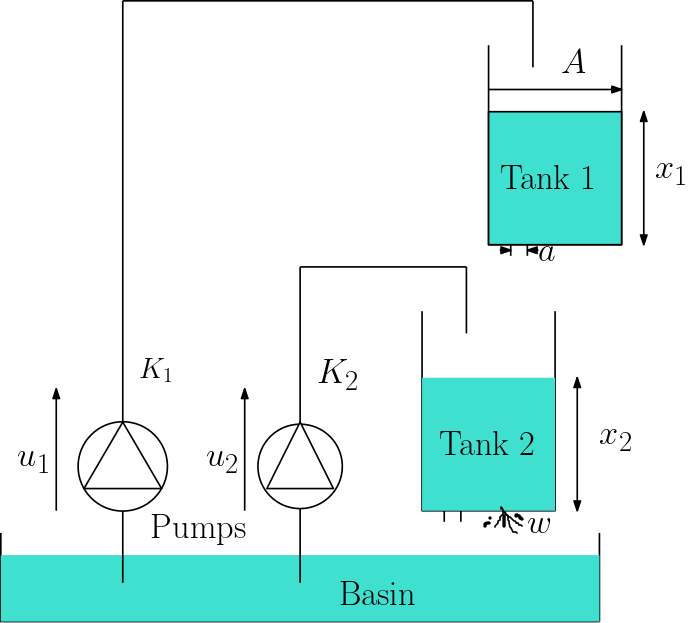}
    \caption{Coupled Tanks setup}
    \label{fig:coupl_tanks}
\end{figure}


$$
\label{eqn:coup_tank}
\begin{array}{rl}
    \left(\begin{array}{l}
\dot{x}_1 \\
\dot{x}_2
\end{array}\right)= & \frac{a \sqrt{2 g}}{A}\left(\begin{array}{cc}
-1 & 0 \\
1 & -1
\end{array}\right)\left(\begin{array}{c}
\sqrt{x_1} \\
\sqrt{x_2}
\end{array}\right)\\
     & +\frac{1}{A}\left(\begin{array}{cc}
K_1 & 0 \\
0 & K_2
\end{array}\right)\left(\begin{array}{l}
u_1 \\
u_2
\end{array}\right)+\frac{1}{A}\left(\begin{array}{l}
0 \\
1
\end{array}\right) d \\
\end{array}
$$

Where $x_1$ and $x_2$ are the water height in tank $1$ and $2$ respectively, $A$ is the cross-sectional of both tanks,  $a$ is the cross-sectional area of the orifice of the two tanks. Both tanks are supplied with two pumps with constants: $K_1$ and $K_2$. $u_1,u_2 \in U=[u_{\min},u_{\max}]$ are voltages of the pump and $d \in D=[d_{\min},d_{\max}]$ represents leakage in tank $2$. One can easily check that the considered system is CSM.

In this example we impose the following safety constraints: $X = \{(x_1, x_2) \mid \,0 \leq  x_1 \leq 30 \mbox{ and } 0 \leq x_2\leq 20\}$. Since the height of the water is always positive, this set is effectively lower-closed. 
We use Algorithm \ref{alg:cap} to compute a robust controlled invariant. The parameters model are
taken from \cite{meyer2015}
and are presented in Table \ref{tab:parameters} 
computed robust controlled invariant set for two different precisions $\epsilon = 1 \mathrm{cm}$ and $\epsilon = 0.5 \mathrm{cm}$ using Algorithm \ref{alg:cap} along with feasible trajectories. In both cases the robust controlled invariant in characterized using feasible trajectories. For the first scenario with $\epsilon = 1 \mathrm{cm}$, the computation time is less than 40 ms. For the scenario with $\epsilon = 0.5 \mathrm{cm}$, the computation time is less than 145 ms. The implementations have been done in Python, on a DELL  Lattitude 5430 using an Intel core i7-1265U. Files for this simulation can be found at this  \href{https://github.com/YoungDevil-glitch/Contro_invariants}{link}.

\begin{table}[h]
    \centering
    \begin{tabular}{|c|c|}
\hline Parameters & Values \\
\hline $\mathrm{A}\left(\mathrm{cm}^2\right)$ & 4.425 \\
\hline $\mathrm{a}\left(\mathrm{cm}^2\right)$ & 0.476 \\
\hline$u_{\min }(\mathrm{V})$ & 0 \\
\hline$u_{\max }(\mathrm{V})$ & 22 \\
\hline$K_1\left(\mathrm{~cm}^3 / \mathrm{V} / \mathrm{s}\right)$ & 4.6 \\
\hline$K_2\left(\mathrm{~cm}^3 / \mathrm{V} / \mathrm{s}\right)$ & 2 \\
\hline$d_{\min }\left(\mathrm{cm}^3 / \mathrm{s}\right)$ & -20 \\
\hline$d_{\max }\left(\mathrm{cm}^3 / \mathrm{s}\right)$ & 0 \\
\hline $\mathrm{g}\left(\mathrm{cm}^2 / \mathrm{s}^2\right)$ & 980 \\
\hline
\end{tabular}
    \caption{Parameters of the coupled tanks system}
    \label{tab:parameters}
\end{table}
\begin{figure}[h]
    \centering
    \includegraphics[width = 0.4 \textwidth]{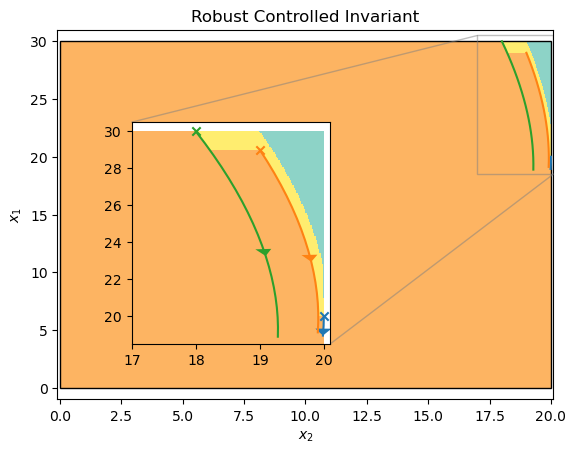}
    \caption{\textcolor{blue}{The  Robust controlled invariant is in brown. The Unsafe set is in teal. The non-explored region of the state constraint set is in yellow. Three feasible trajectories initiated at: $[30,18]$ in green, $[29,19]$ in brown and   $[20,20]$ in light blue, are shown. Feasible trajectories are determined for minimal control inputs.}}
    \label{fig:invariant-1cm}
\end{figure}

\begin{figure}[h]
    \centering
    \includegraphics[width = 0.4 \textwidth]{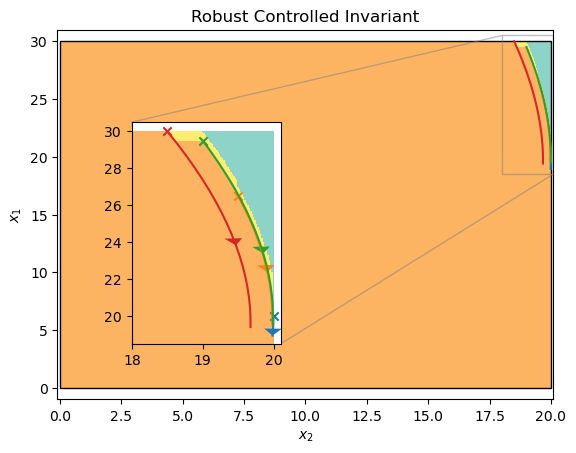}
    \caption{\textcolor{blue}{The  Robust controlled invariant is in brown. The non-explored region of the state constraint set is in yellow. Four feasible trajectories from the following initial conditions: $[30,18.5]$ in red,  $[29.5, 19]$ in green,  $[26.5, 19.5]$ in brown, $[20,20]$ in blue . All feasible trajectories are determined for minimal control inputs.} }
    \label{fig:invariant-05cm}
\end{figure}

\section{Conclusion}
In this study, we have presented characterizations of robust controlled invariants for monotone continuous-time systems. Leveraging these characterizations, we have developed an algorithmic procedure for computing maximal robust controlled invariants. The practical application of our algorithm in an illustrative example highlights the effectiveness of our approach. Moving forward, our future endeavours will delve into the realm of closed-loop robust controlled invariance (while only open-loop controllers have been considered in the current version of the paper), this will be conducted by exploring the potential utilization of tangent cones-based characterizations of invariance.  Additionally, we aim to investigate other types of specifications, particularly those related to stabilization and more general properties described by signal temporal logic formulas.

\bibliography{ifacconf}

\appendix

\section{Monotones Systems Characterization}
\label{sec:charMono}
\begin{proposition}Suppose the system $\Sigma$ in (\ref{dis_sys}) with $x(t) \in \mathcal{X} \subseteq \mathbb{R}^n$, $u(t) \in \mathcal{U} \subseteq \mathbb{R}^m$ and $d(t) \in \mathcal{D} \subseteq \mathbb{R}^p$.
 Suppose that $f$ defines a local Lipschitz vector field in the state space, continuous on the input and disturbance space. The system $\Sigma$ is : 
 \begin{itemize}
     
     \item SM  $\Longleftrightarrow$ for all $x_1,x_2 \in \mathcal{X},\ u \in \mathcal{U}$ and for all $d_1,d_2\in \mathcal{D}^{\mathbb{R}}$, if $x_1 \leq_{\mathcal{X}} x_2$, $d_1 \leq_{\mathcal{D}^{\mathbb{R}}} d_2$ and $x_1^{i} = x_2^{i}$ then $ f_i(x_1,u, d_1)\leq_{\mathcal{X}} f_i(x_2,u,d_2)$ for all $i \in \{ 1, \dots n\} $
     \item CSM $\Longleftrightarrow$ for all $x_1,x_2 \in \mathcal{X},\ u_1,u_2 \in \mathcal{U}$ and for all $d_1,d_2\in \mathcal{D}$, if $x_1 \leq_{\mathcal{X}} x_2$, $u_1 \leq_{\mathcal{U}} u_2$ , $d_1 \leq_{\mathcal{D}} d_2$ and $x_1^{i} = x_2^{i}$ then $ f_i(x_1,u_1, d_1)\leq_{\mathcal{X}} f_i(x_2,u_2,d_2)$ for all $i \in \{ 1, \dots n\} $
 \end{itemize}
\end{proposition}
\begin{remark}
     The notions above can be easily verified via the Kamke-Muller conditions~\cite{smith2008monotone} for continuously differentiable vector fields as follows: The system $\Sigma$ in (\ref{dis_sys}) when $x(t) \in \mathcal{X} \subseteq \mathbb{R}^n$, $u(t) \in \mathcal{U} \subseteq \mathbb{R}^m$ and $d(t) \in \mathcal{D} \subseteq \mathbb{R}^p$ is 
\begin{itemize}

    \item SM  if $\frac{\partial f_i}{\partial x_j} \geq 0$ and $\frac{\partial f_i}{\partial d_h} \geq 0$ for all $i,j \in \{1,2,\ldots,n\} \, i \neq j $ and for all $h\in \{1,2,\ldots,p\}$;
    \item CSM  if $\frac{\partial f_i}{\partial x_j} \geq 0$,  $\frac{\partial f_i}{\partial u_h} \geq 0$ and $\frac{\partial f_i}{\partial d_l} \geq 0$ for all $i,j \in \{1,2,\ldots,n\} \, i \neq j $, for all $h\in \{1,2,\ldots,m\}$ and for all $l \in \{1,2,\ldots,p\}$.
\end{itemize}
where $\geq$ is the usual total order on $\mathbb{R}$.
\end{remark}
\begin{remark}
    Similar results can be defined for other partial orders. For further details, we recommend consulting \cite{angeli2003monotone} for interested readers.
\end{remark}


\section{Proofs}

\textcolor{blue}{\textbf{\underline{Proof of Proposition ~\ref{prop:characterizations_monotone}:}}} 
We will only give proof for $(ii)$, the proofs of the other cases can be derived similarly. 
\begin{itemize}
    \item Suppose that $\Sigma$ is CSM and consider $x_1 \in \Phi(t,\downarrow x, \downarrow \mathbf{u}, \downarrow \mathbf{d})$. Then there exists $x_2 \in \mathcal{X}$, $\mathbf{u}_1 \in U^{\mathbb{R}}$ and $\mathbf{d}_1 \in D^{\mathbb{R}}$  such that $x_2 \leq_{\mathcal{X}} x$,
    $\mathbf{u}_1 \leq_{\mathcal{U}^{\mathbb{R}}} \mathbf{u}$, $\mathbf{d}_1 \leq_{\mathcal{U}^{\mathbb{R}}} \mathbf{d}$ and $x_1 = \Phi(t,x_2, \mathbf{u}_1,\mathbf{d}_1)$. Using the fact that the system is  CSM, we have that $x_1 = \Phi(t,x_2,\mathbf{u}_1,\mathbf{d}_1) \leq_{\mathcal{X}} \Phi(t,x,\mathbf{u},\mathbf{d})$ which implies $x\in \downarrow \Phi(t,x,\mathbf{u},\mathbf{d})$. \textcolor{blue}{Using Lemma 1 from \cite{saoud2024characterization}}, we obtain that  $\Phi( t,\downarrow x,\downarrow \mathbf{u},\downarrow \mathbf{d})\subseteq \downarrow \Phi(t,x,\mathbf{u},\mathbf{d})$. 
    \item Now suppose that for all $x \in \mathcal{X}$, $\mathbf{u} \in \mathcal{U}^{\mathbb{R}}$ and $\mathbf{d} \in \mathcal{D}^{\mathbb{R}}$ we have $ \Phi( t,\downarrow x,\downarrow \mathbf{u},\downarrow \mathbf{d})\subseteq \downarrow \Phi(t,x,\mathbf{u},\mathbf{d})$. Consider $x, x_1 \in \mathcal{X}$, $\mathbf{u}, \mathbf{u}_1 \in \mathcal{U}^{\mathbb{R}}$ and  $\mathbf{d}, \mathbf{d}_1 \in \mathcal{D}^{\mathbb{R}}$ such that $x \leq_{\mathcal{X}} x_1$, $\mathbf{u}\leq_{\mathcal{U}^{\mathbb{R}}} \mathbf{u}_1$ and $\mathbf{d}\leq_{\mathcal{D}^{\mathbb{R}}} \mathbf{d}_1$. Then for all $t \geq 0$, we have that $\Phi(t, x,\mathbf{u}, \mathbf{d}) \in \Phi( t,\downarrow x_1,\downarrow \mathbf{u}_1,\downarrow \mathbf{d}_1)\subseteq \downarrow \Phi(t,x_1,\mathbf{u}_1,\mathbf{d}_1)$, which in turn implies that $\Phi(t, x,\mathbf{u}, \mathbf{d}) \leq_{\mathcal{X}} \Phi(t, x_1,\mathbf{u}_1, \mathbf{d}_1)$ and that the system in CSM.
\end{itemize}
    
\bigskip
  
\textcolor{blue}{\textbf{\underline{Proof of Lemma ~\ref{lem:inv_cont_dist}:}}}
Suppose K is a robust controlled invariant for the system $\Sigma$ and constraint set $(X, U_1, D_1)$. Let $x \in X$, then there exists $\mathbf{u} \in U_1^{\mathbb{R}}$ such that for all $\mathbf{d} \in D_1^{\mathbb{R}}$, 
$\Phi(t, x,\mathbf{u}, \mathbf{d}) \in K$ for all $t > 0$. Since $U_1 \subseteq U_2$ then $\mathbf{u} \in U_2^{\mathbb{R}}$. Since $D_2 \subseteq D_1$, then $D_2^{\mathbb{R}} \subseteq D_1^{\mathbb{R}}$ and one gets for all $\mathbf{d} \in D_2^{\mathbb{R}} \subseteq D_1^{\mathbb{R}}$, $\Phi(t, x,\mathbf{u}, \mathbf{d}) \in K$ for all $t > 0$. Then, K is a robust controlled invariant for constraints  $(X, U_2, D_2)$

\bigskip

\textcolor{blue}{\textbf{\underline{Proof of Proposition~\ref{prop:closedness}:}}}

We provide a proof for each item separately.

\textbf{\underline{Proof of (i):}} 
Since $f$ is Lipschitz we have the existence of $\lambda >0$ such that for all $x_1,x_2 \in X $ for all $u \in U$ and for all $d \in D$ we have
\begin{equation}
\label{eqn:lipschitz}
    \|f(x_1,u,d)-f(x_2,u,d)\| \leq \lambda \|x_1-x_2\|.
\end{equation}
Now consider $x,y \in X$, $\mathbf{u},\mathbf{u}' \in U^{\mathbb{R}}$ and $\mathbf{d},\mathbf{d}' \in D^{\mathbb{R}}$, the following holds for all $t \geq 0$:
\begin{align}
    \|\Phi&(t,x,\mathbf{u},\mathbf{d}) - 
 \Phi(t,y,\mathbf{u}',\mathbf{d}')\| =  \Big{\|} \int\limits_0^t f(\Phi(s,x,\mathbf{u},\mathbf{d}),\mathbf{u},\mathbf{d})ds \nonumber\\ &+x 
   -\Big(\int\limits_0^t f(\Phi(s,y,\mathbf{u}',\mathbf{d}'),\mathbf{u}',\mathbf{d}')ds+y\Big)\Big{\|} \nonumber \\
   \leq&    \int\limits_0^t \Big{\|}f(\Phi(s,x,\mathbf{u},\mathbf{d}),\mathbf{u},\mathbf{d})-f(\Phi(s,y,\mathbf{u}',\mathbf{d}'),\mathbf{u}',\mathbf{d}')\Big{\|}ds \nonumber \\ &+\|x-y\| \nonumber \\ 
     \leq & \int\limits_0^t \Big{\|}f(\Phi(s,x,\mathbf{u},\mathbf{d}),\mathbf{u}',\mathbf{d}')-f(\Phi(s,y,\mathbf{u}',\mathbf{d}'),\mathbf{u}',\mathbf{d}')\Big{\|}ds \nonumber\\ &+   \int\limits_0^t \Big{\|}f(\Phi(s,x,\mathbf{u},\mathbf{d}),\mathbf{u},\mathbf{d})-f(\Phi(s,x,\mathbf{u},\mathbf{d}),\mathbf{u}',\mathbf{d}')\Big{\|}ds \nonumber \\   &+\|x-y\| \nonumber\\  \leq &  \lambda \int\limits_0^t \Big{\|}\Phi(s,x,\mathbf{u},\mathbf{d})-\Phi(s,y,\mathbf{u}',\mathbf{d}')\Big{\|}ds +  \int\limits_0^t h(s)ds \nonumber \\ &+ \|x-y\| 
        \label{eqn:conti_sol}
\end{align}
with $h(s) =\|f(\Phi(s,x,\mathbf{u},\mathbf{d}),\mathbf{u}(s),\mathbf{d}(s))-f(\Phi(s,x,\mathbf{u},\mathbf{d}),\\\mathbf{u}'(s),\mathbf{d}'(s))\|$, where the last inequality comes from (\ref{eqn:lipschitz}). 

To show the result, we proceed by contradiction. Suppose that $K$ is a robust controlled invariant for the system $\Sigma$ and constraint set $(X,U,D)$, and that $\cl(K)$ is not a robust controlled invariant. Then there exists $x\in \cl(K)\setminus K$ such that for all $\mathbf{u} \in U^\mathbb{R}$ there exists $\mathbf{d} \in D^{\mathbb{R}}$ and $t_1 >0$ such that $\Phi(t_1,x,\mathbf{u},\mathbf{d}) \in \mathcal{X}\setminus\cl(K)$. 
Consider a sequence $(x_n)_{n\in \mathbb{N}}$, such that $x_n \in K$ for all $n \in \mathbb{N}$ and $\underset{n \rightarrow +\infty}{x_n\rightarrow x}$. Using the fact that $K$ is robust controlled invariant, we have the existence of a sequence $(\mathbf{u_n})_{n\in \mathbb{N}}$ of functions with $u_n \in U^{\mathbb{R}}$, for all $n \in \mathbb{N}$ and such that $\Phi(t, x_n,\mathbf{u_n}, D) \subseteq K$ for all $n \in \mathbb{N}$ and for all $t\geq 0$. Moreover, using Thychonoff's Theorem (\cite{Etienne3Proof}), one gets that the set $U^{\mathbb{R}}$ is compact. Then, from the sequence $(\mathbf{u_n})_{n\in \mathbb{N}}$, we can extract a sequence which converges pointwise\footnote{The pointwise limite $\mathbf{u} \in U^{\mathbb{R}}$ of the sequence of function $(\mathbf{u_n})_{n\in \mathbb{N}}$ is defined for $t \geq 0$ as $\mathbf{u}(t)=\lim_{n\rightarrow +\infty} \mathbf{u_n}(t)$.} to a function $\mathbf{u} \in U^{\mathbb{R}}$. 

From the assumption that $\cl(K)$ is not a robust controlled invariant, for $x \in cl(K)\setminus K$ chosen above and for $\mathbf{u} \in U^{\mathbb{R}}$ constructed above, we have the existence of a disturbance input $\mathbf{d} \in D^{\mathbb{R}}$ and $t_1 >0$ such that $x' = \Phi(t_1,x,\mathbf{u},\mathbf{d}) \in \mathcal{X}\setminus\cl(K)$. Since $\mathcal{X}\setminus\cl(K)$ is an open set, there exists $\epsilon > 0$ such that 
\begin{equation}
\label{eqn:robust}
    \mathcal{B}_{\epsilon}(x') \subseteq \mathcal{X}\setminus \cl(K).
\end{equation}
Now if we replace by $x_n$, $x$, $\mathbf{u_n}$, $\mathbf{u}$ and $\mathbf{d}$ in (\ref{eqn:conti_sol}) we get for all $t \in [0,t_1]$: 
\begin{align}
\|\Phi(t,x,\mathbf{u},\mathbf{d}) &- 
 \Phi(t,x_n,\mathbf{u}_n,\mathbf{d})\| \leq    \|x-x_n\|+ \int\limits_0^{t} h_n(s)ds \nonumber\\&+\lambda \int\limits_0^{t} \|\Phi(s,x,\mathbf{u},\mathbf{d})-\Phi(s,x_n,\mathbf{u},\mathbf{d})\|ds   
        \label{eqn:conti_sol2}
\end{align} 
where for $s \geq 0$, $ h_n(s) = \|f(\Phi(s,x,\mathbf{u},\mathbf{d}),\mathbf{u},\mathbf{d})-f(\Phi(s,x,\mathbf{u},\mathbf{d}),\mathbf{u}_n,\mathbf{d})\| $. Moreover, using the continuity of the solution with respect to time, one gets that the set $A=\bigcup\limits_{s\in [0,t_1]}\Phi(s,x_0,\mathbf{u},\mathbf{d})$ is compact, then for all $0 \leq s \leq t_1$ and for all $n \in \mathbb{N}$, $h_n(s) \leq 2\sup\limits_{x\in A, u \in U, d \in D}\|(f(x,u,d)\|$, for which the existence follows from the continuity of the map $f$ and the compactness of the set $A \times U \times D$. Then, using the continuity of the map $f$ with respect to the control input, we have that $\underset{n \rightarrow +\infty}{h_n(t)\rightarrow 0}$ for all $t\geq 0$. Hence, from the boundedness and the convergence of the sequence $(\mathbf{u_n})_{n\in \mathbb{N}}$, it follows from the dominated convergence theorem (\cite{royden1988real}) that $ C_n = \int\limits_0^{t} h_n(s)ds \rightarrow 0$. Hence, it follows from (\ref{eqn:conti_sol2}) that for all $t \in [0,t_1]$ that: 

\begin{align*}
    \|\Phi(t,x,\mathbf{u},\mathbf{d}) &- 
 \Phi(t,x_n,\mathbf{u}_n,\mathbf{d})\| \leq    \|x-x_n\|+ c_n(t) \nonumber\\&+\lambda \int\limits_0^{t} \|\Phi(s,x,\mathbf{u},\mathbf{d})-\Phi(s,x_n,\mathbf{u},\mathbf{d})\|ds   
        \label{eqn:conti_sol3}
\end{align*}

 Using The comparison Lemma (\cite{khalil2002nonlinear}), we have that for all $t \in [0,t_1]$
 \begin{equation}
 \label{eqn:conti_sol4}
    \begin{array}{rl}
        \|\Phi(t,x_0,\mathbf{u},\mathbf{d}) - 
 \Phi(t,x_n,\mathbf{u}_n,\mathbf{d})\| \leq &   (\|x-x_n\|+C_n)e^{\lambda t}  
        \end{array}
 \end{equation}
Moreover, since $\|x-x_n\|+C_n \rightarrow 0$, there exists $N \in \mathbb{N}$ such that for all $n \geq N$, we have $\|x-x_n\|+C_n < \epsilon e^{-\lambda t_1}$. Hence, from (\ref{eqn:conti_sol4}) for $t=t_1$, one gets that $\|\Phi(t_1,x_0,\mathbf{u},\mathbf{d}) - 
 \Phi(t_1,x_n,\mathbf{u}_n,\mathbf{d})\| < \epsilon$. The later implies from (\ref{eqn:robust}) that $\Phi(t_1,x_n,\mathbf{u}_n,\mathbf{d}) \in \mathcal{B}_{\epsilon}(x') \subseteq \mathcal{X}\setminus \cl(K)$, which contradits the fact that $\Phi(t_1,x_n,\mathbf{u}_n,\mathbf{d}) \in K \subseteq \cl(K)$. Hence, the set $\cl(K)$ is then a robust controlled invariant for the system $\Sigma$ and constraint set $(X,U,D)$.  
 
\textbf{\underline{Proof of (ii):}} Let $K$ be the maximal robust controlled invariant for the system $\Sigma$ and constraint set $(X,U,D)$. Since $X$ is closed ,$\cl(K) \subseteq X$ . Using (i) $\cl(K)$ is also a robust controlled invariant for the system $\Sigma$ and constraint set $(X, U,D)$. Since  $K$ is maximal, $\cl(K)=K$. Hence, $K$ is closed.

\bigskip
  
    \textcolor{blue}{\textbf{\underline{Proof of Theorem~\ref{thm1}:}}}

We provide proof for each item separately.

\textbf{\underline{proof of (i):}} Let $K$ be a robust controlled invariant for the system $\Sigma$ and constraint set $(X,U,D)$. Consider the set $H =\downarrow K$ and let us show that the set $H$ is a robust controlled invariant for the system $\Sigma$ and constraint set $(X, U,D)$. Choose any $x\in H$ then there exist a $x_1 \in K$ such that $x \leq_{\mathcal{X}} x_1$. Since $K$ is a robust controlled invariant, there exists $\mathbf{u} \in U^{\mathbb{R}}$ such that  for all $t \geq 0 $ $\Phi(t, x_1 , \mathbf{u}, D) \subseteq K$. Since $\Sigma$ is SM, for any $d \in D^{\mathbb{R}}$, 
for all $t\geq 0$ we have that $\Phi(t, x, \mathbf{u},d) \leq_{\mathcal{X}} \Phi(t, x_1, \mathbf{u},d)$. Hence, from (i) in Proposition \ref{prop:characterizations_monotone}, we get that $\Phi(t, x, \mathbf{u},D) \subseteq \downarrow \Phi(t, x_1, \mathbf{u},  D) \subseteq \downarrow K = H$ and  $H$ is a robust controlled invariant for the system $\Sigma$ and constraint set $(X,U,D)$.

In the following , $K$ is the maximal robust controlled invariant for the system $\Sigma$ and constraint set $(X, U,D)$. 
 \textbf{\underline{proof of (ii):}} Consider the set $H =\downarrow K$. First, we have from (i) that the set $H$ is a robust controlled invariant for the system $\Sigma$ and constraint set $(X, U, D)$. Moreover, since $K$ is the maximal robust controlled invariant for the system $\Sigma$ and constraint set $(X, U, D)$, one has $H=\downarrow K \subseteq K$. Finally, using the fact that $K \subseteq \downarrow K=H$, one gets $K =\downarrow K$, implying that $K$ is a lower closed set.

\textbf{\underline{proof of (iii):}} Consider $\overline{K}$ be the maximal robust controlled invariant for the system $\Sigma$ and the and constraint set $(X,U,D_{\max})$. First, since $D_{\max} \subseteq D$, we have from Lemma~\ref{lem:inv_cont_dist} that $K \subseteq \overline{K}$. To show that $\overline{K} \subseteq K$, and from the maximality of the set $K$ it is sufficient to show that the set $\overline{K}$ is a controlled invariant of the system $\Sigma$ and constraint set $(X, U,D)$. 
Consider $x \in \overline{K}$, we have the existence of a control $\mathbf{u}: \mathbb{R}_{\geq 0} \rightarrow U$ such that  $\Phi(t,x,\mathbf{u},D_{\max}) \subseteq \overline{K}$ for all $t\geq 0$. Moreover, since the set of disturbance inputs $D$ is closed and bounded above and using the fact that $\Sigma$ is SM, one has from (iii) in Proposition~\ref{prop:characterizations_monotone} that $\Phi(t,x,\mathbf{u},D) \subseteq \Phi(t,x,\mathbf{u},\downarrow D_{\max}) \subseteq \downarrow \Phi(t,x,\mathbf{u},D_{\max})$ for all $t \geq 0$. Hence, one gets that $\Phi(t,x,D) \subseteq  \downarrow \overline{K}= \overline{K}$, for all $t \geq 0$, where the last equality follows from (i). Hence $\overline{K} \subseteq K$ and (iii) holds.

\textbf{\underline{proof of (iv):}} Consider $\underline{K}$ be the maximal robust controlled invariant for the system $\Sigma$ and the constraint set $(X,U_{\min}, D)$. First, since $U_{\min} \subseteq U$, we have from Lemma~\ref{lem:inv_cont_dist} that $\underline{K} \subseteq K$. To show that $K \subseteq \underline{K}$, from the maximality of the set $\underline{K}$, it is sufficient to show that the set $K$ is a controlled invariant of the system $\Sigma$ and constraint set $(X, U_{\min}, D)$.
Consider $x \in K$, we have the existence of $\mathbf{u}: \mathbb{R}_{\geq 0} \rightarrow U$ such that   $\Phi(t,x,\mathbf{u}, D) \subseteq K$ for all $t \geq 0$. Moreover, since the set of control inputs $U$ is closed and bounded below we have the existence of $\underline{\mathbf{u}}:  \mathbb{R}_{\geq 0} \rightarrow  U_{\min}$ such that $\underline{\mathbf{u}} \leq_{\mathcal{U}^{\mathbb{R}}} \mathbf{u}$. Since the system $\Sigma$ is CSM, one has from (ii) in Proposition~\ref{prop:characterizations_monotone} that  $ \Phi(t,x,\underline{\mathbf{u}}, D) \subseteq  \downarrow \Phi(t,x,\mathbf{u}, D)$ for all $t\geq 0$. Hence $\Phi(t,x,\underline{\mathbf{u}}, D)  \subseteq \downarrow K = K \subseteq \underline{K}$ for all $t \geq 0$ and (iv) holds.

\textbf{\underline{proof of (v):}} 
The proof is a direct conclusion from (iii), (iv) and the fact that any CSM system is a SM system.

\medskip
  
    \textcolor{blue}{\textbf{\underline{Proof of Proposition~\ref{prop:charac2}:}}}

We provide proof for each item separately.

\textbf{\underline{proof of (i):}} First, it can be easily seen that if the set $K$ is a robust controlled invariant for the system $\Sigma$ and constraint set $(X, U, D)$ and using the fact that $\max(K) \subseteq K$ and $D_{\max} \subseteq D$, one gets the required result. Now assume that (\ref{eqn:prop_charact1}) holds and let us show that (\ref{eqn:prop_charact}) holds.  
Consider $x \in K$, we have the existence of $x' \in \max(K)$ such that $x \leq_{\mathcal{X}} x'$. Then, from (\ref{eqn:prop_charact1}), one has the existence of $\mathbf{u} \in U^{\mathbb{R}}$ such that $\Phi(t, x',\mathbf{u},D_{\max}) \subseteq K$ for all $t \geq 0$. Hence, one gets that $\Phi(t, x,\mathbf{u},D) \subseteq \Phi(t, x',\mathbf{u},D) \subseteq \Phi(t, x',\mathbf{u},\downarrow D_{\max}) \subseteq \downarrow \Phi(t, x',\mathbf{u},D_{\max}) \subseteq \downarrow K \subseteq K$.  The first inclusion comes from (i) in Proposition~\ref{prop:characterizations_monotone}, the second from the  fact that $D \subseteq \downarrow D_{\max}$, the third inclusion comes from (iii) in Proposition~\ref{prop:characterizations_monotone} and the last inclusion comes from the fact that $K$ is lower closed. Hence, condition (\ref{eqn:prop_charact}) holds. Then $K$ is a robust controlled invariant for the system $\Sigma$ and constraint set $(X, U, D)$. 

\textbf{\underline{proof of (ii):}} From (i) and since the system $\Sigma$ is CSM, hence SM, to show (ii), it is sufficient to show the equivalence between conditions (\ref{eqn:prop_charact1}) and (\ref{eqn:prop_charact2}). Since $U_{\min} \subseteq U$, one gets directly that (\ref{eqn:prop_charact2}) implies (\ref{eqn:prop_charact1}). Let us show the converse result, consider $x \in \max(K)$, from (\ref{eqn:prop_charact1}) one has the existence of $\mathbf{u} \in U^{\mathbb{R}}$ such that $\Phi(t, x,\mathbf{u},D_{\max}) \subseteq K$ for all $t \geq 0$. Choose any  $\underline{\mathbf{u}}:  \mathbb{R}_{\geq 0} \rightarrow U_{\min}$ such that $\underline{\mathbf{u}} \leq_{\mathcal{U}^{\mathbb{R}}} \mathbf{u}$. Then we that for all $t \geq 0$ that $\Phi(t, x,\underline{\mathbf{u}}, D_{\max}) \subseteq \Phi(t, x, \downarrow \mathbf{u}, D_{\max}) \subseteq \downarrow \Phi(t, x,\mathbf{u}, D_{\max}) \subseteq \downarrow K \subseteq K$, where the second inclusion comes from (ii) in Proposition~\ref{prop:characterizations_monotone} and the last inclusion comes from the fact that $K$ is lower closed. Hence, (\ref{eqn:prop_charact}) hold.

\bigskip

 \textcolor{blue}{\textbf{\underline{Proof of Proposition~\ref{prop:feas}:}}}

\textbf{\underline{Necessary condition:}} From the feasibility of $x_0$ w.r.t the constraint set $(X,U,D)$ we have the existence of a control input $\mathbf{u}:\mathbb{R}_{\geq 0} \rightarrow U$ and $T>0 $ such that (\ref{eqn:feas1o}) and (\ref{eqn:feas2o}) hold. Using (\ref{eqn:feas1o}) and the fact that $D_{\max} \subseteq D$, we have that $\Phi(t,x_0,\mathbf{u},D_{\max}) \subseteq \Phi(t,x_0,\mathbf{u},D) \subseteq X$ for all $0 \leq t< T$. Let us show that the second condition holds. We have
 \begin{align*}
   \Phi(T,x_0,\mathbf{u},D_{\max}) &\subseteq \Phi(T,x_0,\mathbf{u},D)\\ &\subseteq \downarrow \bigcup\limits_{0 \leq t < T} \Phi(t,x_0,\mathbf{u},D)\\ &\subseteq \downarrow \bigcup\limits_{0 \leq t <T} \Phi(t,x_0,\mathbf{u},D_{\max})
 \end{align*}
where the first inclusion follows from the fact that $D_{\max} \subseteq D$, the second inclusion comes from (\ref{eqn:feas2o}) and the last inclusion comes from Lemma~\ref{lem:DSM_feas}.

\textbf{\underline{Sufficient condition:}} From the feasibility of $x_0$ w.r.t the constraint set $(X,U,D_{\max})$ we have the existence of a control input $\mathbf{u}:\mathbb{N}_{\geq 0} \rightarrow U$ and $T>0 $ such that the following conditions are satisfied
\begin{equation}
    \label{eqn:feaspr1}
    \Phi(t,x_0,\mathbf{u},D_{\max}) \subseteq X, \quad \forall~ 0<t<T
\end{equation}
and 
\begin{equation}
    \label{eqn:feaspr2}
    \Phi(T,x_0,\mathbf{u},D_{\max}) \subseteq \downarrow \bigcup\limits_{0 \leq t <T} \Phi(t,x_0,\mathbf{u},D_{\max})
\end{equation}
First, we have $\Phi(t,x_0,\mathbf{u}, D) \subseteq  \Phi(t,x_0,\mathbf{u},\downarrow D_{\max}) \subseteq \downarrow  \Phi(t,x_0,\mathbf{u}, D_{\max}) \subseteq \downarrow X=X$, for all $0 \leq t <T$, where the first inequality comes from Lemma~\ref{lem:DSM_feas}, the second inequality follows from (\ref{eqn:feaspr1}) and the last inequality comes from the lower closedness of the set $X$. To show that (\ref{eqn:feas2o}) holds, we have the following 
\begin{align*}
    \Phi(T,x_0,\mathbf{u},D) &\subseteq \downarrow \Phi(T,x_0,\mathbf{u},D_{\max})\\ &\subseteq  \downarrow \bigcup\limits_{0 \leq t<T } \Phi(t,x_0,\mathbf{u},D_{\max})\\ &\subseteq \downarrow \bigcup\limits_{0 \leq t <T} \Phi(t,x_0,\mathbf{u},D)
    \end{align*}
where the first inclusion comes from Lemma~\ref{lem:DSM_feas}, the second inclusion comes from the fact that $x_0$ is feasible w.r.t the constraint set $(X,U,D_{\max})$ and the last inclusion follows from the fact that $D_{\max} \subseteq D$

\bigskip

    \textcolor{blue}{\textbf{\underline{Proof of Theorem ~\ref{prop:feas_inv}:}}}
 Assume that $x_0$ is feasible w.r.t the constraint set $(X,U,D)$. Since the system is SM, from Proposition \ref{prop:feas}, $x_0$ is feasible w.r.t the constraint set $(X,U,D_{max})$ where $D_{max} = \{d_{\max}\}$.  Hence, there exist a controller $\mathbf{u}:\mathbb{R}_{\geq 0}\rightarrow U$ and $T>0$ such that conditions (\ref{eqn:feas1o}) and (\ref{eqn:feas2o}) are satisfied. Let $K = \downarrow \bigcup\limits_{0 \leq t \leq T} \Phi (t, x_0,\mathbf{u}, D) = \downarrow \bigcup\limits_{0 \leq t\leq T} \Phi (t, x_0,\mathbf{u},D_{\max}) $, where the last equality comes from Lemma \ref{lem:DSM_feas}. Consider $x \in K$, we have the following two cases. 
 \begin{itemize}
     \item \underline{Case $1$, $x = x_0$:} Equation (\ref{eqn:feas2o}) can be rewritten as $$\Phi(T, x_0,\mathbf{u},\overline{\mathbf{d}}) \in  \downarrow \bigcup\limits_{0 \leq t<T} \Phi (t, x_0,\mathbf{u},\overline{\mathbf{d}}).$$ With $\overline{\mathbf{d}}:\mathbb{R}_{\geq 0} \rightarrow D$ is such that $\overline{\mathbf{d}}(t)=d_{\max}$ for all $t \geq 0$. Then there exists $0 \leq \delta < T$ such that 
     \begin{equation}
     \label{eqn:feas_inv}
     \Phi(T, x_0,\mathbf{u},\overline{\mathbf{d}}) \leq_{\mathcal{X}} \Phi (T-\delta, x_0,\mathbf{u},\overline{\mathbf{d}}).
     \end{equation}
     We construct the following control input $\overline{\mathbf{u}} : \mathbb{R}_{\geq 0} \rightarrow U$: 
     \begin{equation}
     \label{eqn:input}
        \overline{\mathbf{u}} = \begin{cases}
         \mathbf{u}(t) \mbox{ if } 0 \leq t \leq T\\
         \mathbf{u}(t - \bigg(\bigg\lceil\frac{t-T}{\delta}\bigg\rceil\bigg)\delta) ) \mbox{ if } t > T
     \end{cases}
     \end{equation}
     For all $t > T$, we have that $t-T <\lceil\frac{t-T}{\delta}\rceil \delta \leq t-T+ \delta$, which implies that $T- \delta < t-\lceil\frac{t-T}{\delta}\rceil \delta \leq T$, and that the control input $\overline{\mathbf{u}} : \mathbb{R}_{\geq 0} \rightarrow \mathcal{U}$ is well defined.
     
     Consider $T<t \leq T+\delta$,
     we have the following: 
     \begin{equation}
         \label{eqn:feas_ext}
         \begin{array}{rl}
             \Phi(t, x_0,\overline{\mathbf{u}},\overline{\mathbf{d}})   = &  \Phi(t-T, \Phi(T, x_0,\mathbf{u},\overline{\mathbf{d}}),\overline{\mathbf{u}}_1,\overline{\mathbf{d}}) \\
              \leq_{\mathcal{X}}  & \Phi(t-T, \Phi (T-\delta, x_0,\mathbf{u},\overline{\mathbf{d}}),\overline{\mathbf{u}}_1,\overline{\mathbf{d}})\\
              \leq_{\mathcal{X}} & \Phi (t-\delta, x_0,\mathbf{u},\overline{\mathbf{d}})
         \end{array}
     \end{equation}
     with $\overline{\mathbf{u}}_1(t) = \overline{\mathbf{u}}(t+T) $, where the equality comes from the definition of $\mathbf{u}_1$ and the first inequality follows from (\ref{eqn:feas_inv}). Then, one gets from definition of the set $K$ and the fact that $K$ is lower closed set that for all  $T<t \leq T+\delta$, $\Phi(t, x_0,\overline{\mathbf{u}},\overline{\mathbf{d}}) \in  K$ and $\Phi(T+\delta, x_0,\overline{\mathbf{u}},\overline{\mathbf{d}}) \leq_{\mathcal{X}} \Phi (T, x_0,\mathbf{u},\overline{\mathbf{d}}) \leq_{\mathcal{X}} \Phi (T-\delta, x_0,\mathbf{u},\overline{\mathbf{d}})$.

     Now, suppose that there exists $n> 1$ and  $t \in (T+n\delta, T+(n+1)\delta]$ such that $$\lnot [\Phi(t, x_0,\overline{\mathbf{u}},\overline{\mathbf{d}}) \leq_{\mathcal{X}}  \Phi (t -(T+n\delta)+T-\delta, x_0,\mathbf{u},\overline{\mathbf{d}})].$$ 
     
     Let $n_0 >1$ be defined as follows: $n_0 = \max\{n \in \mathbb{N}| \text{ s.t for all }  i\in \{0,1,\ldots,n_0-1\}  \text{ and for all } t \in (T+i\delta, T+(i+1)\delta] \text{ we have } \Phi(t, x_0,\overline{\mathbf{u}},\overline{\mathbf{d}}) \leq_{\mathcal{X}}  \Phi ((t-T-i\delta)+T-\delta, x_0,\mathbf{u},\overline{\mathbf{d}}) \}$. 
     
     We then have from definition of $n_0$ that $\Phi(T+n_0\delta, x_0,\overline{\mathbf{u}},\overline{\mathbf{d}}) =  \Phi(T+(n_0-1)\delta + \delta, x_0,\overline{\mathbf{u}},\overline{\mathbf{d}})\leq_{\mathcal{X}}  \Phi (T-\delta+\delta, x_0,\mathbf{u},\overline{\mathbf{d}}) \leq_{\mathcal{X}}  \Phi (T-\delta, x_0,\mathbf{u},\overline{\mathbf{d}})$. Then, Equation (\ref{eqn:feas_ext}) can be rewritten for $T_{n_0}= T+n_0\delta< t  \leq T+(n_0+1)\delta = T_{n_0+1}$ as: 
     \begin{equation}
         \label{eqn:feas_ext2}
         \begin{array}{rl}
             \Phi(t, x_0,\overline{\mathbf{u}},\overline{\mathbf{d}})   = &  \Phi(t-T_{n_0}, \Phi(T_{n_0}, x_0,\mathbf{u},\overline{\mathbf{d}}),\overline{\mathbf{u}}_2,\overline{\mathbf{d}}) \\
              \leq_{\mathcal{X}}  & \Phi(t-T_{n_0}, \Phi (T-\delta, x_0,\mathbf{u},\overline{\mathbf{d}}),\overline{\mathbf{u}}_2,\overline{\mathbf{d}})\\
              \leq_{\mathcal{X}} & \Phi (t-T_{n_0}+T-\delta, x_0,\mathbf{u},\overline{\mathbf{d}})
         \end{array}
     \end{equation}
     with $\overline{\mathbf{u}}_2(t) = \overline{\mathbf{u}}(t+T+n_0\delta) $.  This contradicts the maximality of $n_0$. Then for all $n\in \mathbb{N}$, Then, we have for all $n\in \mathbb{N}$ and for all $t \in (T+n\delta, T+(n+1)\delta]$ that \begin{align*}
        \Phi(t, x_0,\overline{\mathbf{u}},\overline{\mathbf{d}}) &\leq_{\mathcal{X}}  \Phi (t -(T+n\delta)+T-\delta, x_0,\mathbf{u},\overline{\mathbf{d}}) \\  &\in \downarrow \bigcup\limits_{T- \delta\leq t \leq T}  \Phi (t, x_0,\mathbf{u},\overline{\mathbf{d}}).\end{align*} Using the result from Equations (\ref{eqn:feas_ext}) and (\ref{eqn:feas_ext2}) and the feasibility of $x_0$, we have that for all $t\geq 0$, $\Phi(t, x_0,\overline{\mathbf{u}},\overline{\mathbf{d}}) \in K$ 
     \item \underline{Case $2$, $x \neq x_0$:} we have the existence of $t_x \in [0,T)$ such that $x \leq_{\mathcal{X}} \Phi(t_x,x_0,\mathbf{u},\overline{\mathbf{d}}) = \Phi(t_x,x_0,\overline{\mathbf{u}},\overline{\mathbf{d}})$, where the equality follows from the construction of $\overline{\textbf{u}}$ in (\ref{eqn:input}). Consider the input signal $\overline{\mathbf{u}}_3 \in U^{\mathbb{R}}$ defined for $t \geq 0$ as $\overline{\mathbf{u}}_3(t) = \overline{\mathbf{u}}(t+t_x)$. Using the fact that the system $\Sigma$ is SM we have  that 
 \begin{equation}
 \label{eqn:feas_inv_final}
     \begin{array}{rl}
        \Phi(t,x,\overline{\mathbf{u}}_3,\mathbf{d}) \leq_\mathcal{X} &  \Phi(t,\Phi(t_x,x_0,\overline{\mathbf{u}},\overline{\mathbf{d}}),\overline{\mathbf{u}}_3,\overline{\mathbf{d}}) \\
         \leq_\mathcal{X} & \Phi(t+t_x,x_0,\overline{\mathbf{u}},\overline{\mathbf{d}})
     \end{array}
 \end{equation}
 By definition of $\overline{\mathbf{u}}$ and the lower-closedness of the $K$,  one gets that $\Phi(t,x,\overline{\mathbf{u}}_3,\mathbf{d}) \in K$, for all $t \geq 0$.
\end{itemize}
 Hence, using (ii) in Proposition \ref{prop:charac2} we conclude that $K$ is a robust controlled invariant for system $\Sigma$ and constraint $(X,U,D_{\max})$.

\bigskip 

\textcolor{blue}{\textbf{\underline{Proof of Proposition~\ref{prop:feas_CSM}:}}}
    Consider $x_0 \in X$, first since $U_{\min} \subseteq U$ one directly has that $x_0$ is feasible w.r.t the constraint set $(X,U,D)$ if $x_0$ is feasible w.r.t the constraint set $(X,U_{\min},D)$. Let us show the second implication. 
Since $x_0$ is feasible, there exist $\mathbf{u} : \mathbb{R}_{\geq 0} \rightarrow U$ and $T >0$ such that : 
\begin{equation}
    \Phi(t,x_0,\mathbf{u},D) \subseteq X, \quad \forall~ 0<t<T 
\end{equation}
and 
\begin{equation}
 \Phi(T,x_0,\mathbf{u},D) \subseteq \downarrow \bigcup\limits_{0 \leq t <T} \Phi(t,x_0,\mathbf{u},D).
\end{equation}
Using the result, from Proposition \ref{prop:feas},  $x_0$ is feasible w.r.t the constraint set $(X,U,D_{max})$ where $D_{max} = \{d_{\max}\}$, we just have to show that $x_0$ is feasible with respect to constraint set   $(X,U_{min},D_{max})$. Consider $\underline{\mathbf{u}} : \mathbb{R}_{\geq 0} \rightarrow U_{\min}$ such that $\underline{\mathbf{u}} \leq_{\mathcal{U}^{\mathbb{R}}} \mathbf{u} $. Since $\Sigma$ is CSM, we have from (iv) in Proposition \ref{prop:characterizations_monotone} that $ \Phi(T,x_0,\underline{\mathbf{u}}, D_{max}) \subseteq  \Phi(T,x_0,\mathbf{u},D_{max}) \subseteq \downarrow \bigcup\limits_{0 \leq t <T} \Phi(t,x_0,\mathbf{u},D_{max})$. Using the fact that $D_{max} = \{d_{max}\}$, we have that $D_{max}^{\mathbb{R}} =\{\overline{\mathbf{d}}: t \mapsto d_{max}\}$. Which implies the existence of $0 <t_1 <T$ such that: 
 \begin{equation}
     \label{eqn:pr22}
     \Phi(T,x_0,\underline{\mathbf{u}},\overline{\mathbf{d}}) \leq_{\mathcal{X}}\Phi(T,x_0,\mathbf{u},\overline{\mathbf{d}}) \in \downarrow \Phi(t_1,x_0,\mathbf{u},\overline{\mathbf{d}})
     \end{equation}.
Let $\epsilon :=  \Phi(t_1,x_0,\mathbf{u},\overline{\mathbf{d}}) - \Phi(t_1,x_0,\underline{\mathbf{u}},\overline{\mathbf{d}}).$
\begin{equation*}
    \begin{array}{rl}
       \Phi(T,x_0,\underline{\mathbf{u}},\overline{\mathbf{d}})  \leq_{\mathcal{X}}  & \Phi(T-t_1 ,\Phi(t_1,x_0,\underline{\mathbf{u}},\overline{\mathbf{d}}),\underline{\mathbf{u}}',\overline{\mathbf{d}})\\
       \leq_{\mathcal{X}}  &  \Phi(T-t_1 ,\Phi(t_1,x_0,\underline{\mathbf{u}},\overline{\mathbf{d}}),\mathbf{u}',\overline{\mathbf{d}})\\
         \leq_{\mathcal{X}}& \Phi(T-t_1 ,\Phi(t_1,x_0,\mathbf{u},\overline{\mathbf{d}}),\mathbf{u}',\overline{\mathbf{d}})- \epsilon\\
         \leq_{\mathcal{X}}& \Phi(T,x_0,\mathbf{u},\overline{\mathbf{d}})- \epsilon\\
         \leq_{\mathcal{X}}& \Phi(t_1,x_0,\mathbf{u},\overline{\mathbf{d}}) - \epsilon\\
         \leq_{\mathcal{X}}& \Phi(t_1,x_0,\underline{\mathbf{u}},\overline{\mathbf{d}})
    \end{array}
\end{equation*}
where $\mathbf{u}'(t) = \mathbf{u}(t+t_1)$ , $\underline{\mathbf{u}}'(t) = \underline{\mathbf{u}}(t+t_1)$. 
where the second inequality follow from the application of (\ref{eqn:feas_CSM_C}) to the term $\epsilon$ for $t = T-t_1$ and equation (\ref{eqn:pr22}). The third inequality comes from the fact that the system $\Sigma$ is CSM. The last equality comes from the definition of $\varepsilon$. Hence $x \leq_{\mathcal{X}} \Phi(t_1,x_0,\underline{\mathbf{u}},\Bar{\mathbf{d}})$. Hence, condition (\ref{eqn:feas2o}) is satisfied and $x_0$ is feasible w.r.t the constraint set $(X,U_{\min},D_{max})$. Proposition \ref{prop:feas},  $x_0$ is feasible w.r.t the constraint set $(X,U_{min},D)$. 

\bigskip

\textcolor{blue}{\textbf{\underline{Proof of Theorem~\ref{thm:stric_feas}:}}}
      Since $f$ is uniformly continuous with respect to $u$ and $d$. Then  for all $\beta > 0$, there exists $\alpha >0$ such that for all $x\in X$, for all $u,u_1 \in U$ and for all $d, d_1 \in D$, the following holds
\begin{equation}
\label{eqn:conti}
 \|u-u_1\|+\|d-d_1\| \leq \alpha \implies \|f(x,u,d) - f(x,u_1,d_1)\| \leq \beta.
 \end{equation}
Let $\epsilon = \min(\epsilon_T, \gamma)$,  $\beta < \epsilon $, $x \in X$ such that $\|x - x_0\| \leq \beta$ and consider $\mathbf{d},\mathbf{d}_1 \in D^{\mathbb{R}}$ satisfying $\|\mathbf{d}_1 - \mathbf{d}\|_{\infty} \leq \alpha$. To simplify notation, let us define $\mathbf{x}_{\mathbf{d}_1}(t) = \Phi(t,x, \mathbf{u},\mathbf{d}_1) $ and $\mathbf{x}_{0,\mathbf{d}}(t) = \Phi(t,x_0, \mathbf{u},\mathbf{d})$. 
We then have for all $t\geq 0$: 
\begin{equation*}
    \begin{array}{rl}
       \scriptstyle{ \|\mathbf{x}_{\mathbf{d}_1}(t)- \mathbf{x}_{0,\mathbf{d}}(t) \| \leq}  & \scriptstyle{ \|x- x_0 \| + \int\limits_0^t\|f(\mathbf{x}_{\mathbf{d}_1}(s),\mathbf{u},\mathbf{d}_1) - f(\mathbf{x}_{0,\mathbf{d}}(s),\mathbf{u},\mathbf{d})\| ds }  \\
         \scriptstyle{\leq} & \scriptstyle{ \|x- x_0 \| + \int\limits_0^t\|f(\mathbf{x}_{\mathbf{d}_1}(s),\mathbf{u},\mathbf{d}_1) - f(\mathbf{x}_{0,\mathbf{d}}(s),\mathbf{u},\mathbf{d}_1)\| ds} \\ & \scriptstyle{+ \int\limits_0^t \beta ds}  \\
          \scriptstyle{\leq} & \scriptstyle{ \|x- x_0 \| + \lambda \int\limits_0^t\|\mathbf{x}_{\mathbf{d}_1}(s) - \mathbf{x}_{0,\mathbf{d}}(s)\| ds + \beta t }  \\
          
    \end{array}
\end{equation*}
The first inequality comes from the triangular inequality. The second inequality comes from (\ref{eqn:conti}). The last inequality comes from the Lipschitzness of $f$. Using the Gronwall-Bellman inequality, one gets for all $t \geq 0$: 
$$ \begin{array}{rl}
\scriptstyle{ \|\mathbf{x}_{\mathbf{d}_1}(t)- \mathbf{x}_{0,\mathbf{d}}(t) \| \leq}  & \scriptstyle{ (\|x- x_0 \| + \beta t ) + \int\limits_0^t\lambda (\|x- x_0 \| + \beta s )e^{\lambda(t-s)} ds }  \\
  \scriptstyle{  \leq}  &   \scriptstyle{\|x- x_0 \|e^{\lambda t} +   \frac{\beta}{\lambda} (e^{\lambda t} -1  )} \\
      \scriptstyle{ \leq} & \scriptstyle{ \beta\left(1+ \frac{1}{\lambda} \right) e^{\lambda t}}
\end{array}$$
Let us choose $\beta = \epsilon \frac{\lambda e^{-\lambda T}}{1+\lambda} $. This choice ensure that for all $t \in [0,T]$ we have 
\begin{equation}
\label{eqn:Thm8_pr}
    \|\mathbf{x}_{\mathbf{d}_1}(t)- \mathbf{x}_{0,\mathbf{d}}(t) \| \leq \epsilon.
    \end{equation}
 Now consider $x \in \mathcal{B}_{\beta}(x_0)$, we have from  Equation (\ref{eqn:Thm8_pr}), that for all $0 \leq  t < T$  
$$\Phi(t,x,\mathbf{u},D) \subseteq \mathcal{B}_{\gamma}(\Phi(t,x_0,\mathbf{u},D)) \subseteq X.$$ 
Moreover, for any $x_1 \in \mathcal{B}_{\beta}(x_0) \bigcap (\uparrow x_0)$, we have from  (\ref{eqn:Thm8_pr}) and (\ref{eqn:thm}) that: 
$$\begin{array}{rl}
     \Phi(T,x_1,\mathbf{u},D) \subseteq & 
 \mathcal{B}_{\epsilon_T}(\Phi(T,x_0,\mathbf{u},D))\\
     \subseteq & \downarrow \bigcup\limits_{0 \leq t <T} \Phi(t,x_0,\mathbf{u},D)  \\
     \subseteq & \downarrow \bigcup\limits_{0 \leq t <T} \Phi(t,x_1,\mathbf{u},D)
\end{array}$$
where the last inclusion comes from the fact that the system is $SM$, which concludes the proof.


\end{document}